\documentclass{mfa}
\usepackage[utf8]{inputenc}
\usepackage{graphicx}
\usepackage{booktabs}
\usepackage{color}
\usepackage{hyperref}

\title[Analyzing the PACT]{Analysis of the Pennsylvania Additive Classification Tool:  Biases and Important Features}
\author[S. Dhar]{Swarup Dhar}
\address{Departments of the Mathematics and Computer Science Enginerring, Bucknell University}
\author[V. Massaro]{Vanessa Massaro}
\address{Department of Geography, Bucknell University}
\author[D. Mir]{Darakhshan J. Mir}
\address{Department of Computer Science, Bucknell University}
\author[N. Ryan]{Nathan C. Ryan}
\address{Department of Mathematics, Bucknell University}


\begin{document}
\begin{abstract}
    The Pennsylvania Additive Classification Tool (PACT) is a carceral algorithm used by the Pennsylvania Department of Corrections in order to determine the security level for an incarcerated person in the state's prison system.  For a newly incarcerated person it is used in their initial classification.  The initial classification can be overridden both for discretionary and administrative reasons.  An incarcerated person is reclassified annually using a variant of the PACT and this reclassification can be overridden, too, and for similar reasons.  In this paper, for each of these four processes (the two classifications and their corresponding overrides), we develop several logistic models, both binary and multinomial, to replicate these processes with high accuracy.  By examining these models, we both identify which features are most important in the model and quantify and describe biases that exist in the PACT, its overrides, and its use in reclassification.  Because the details of how the PACT operates have been redacted from public documents, it is important to know how it works and what disparate impact it might have on different incarcerated people.
\end{abstract} 

\maketitle

\section{Introduction}

For more than a century so-called actuarial methods have been used in carceral systems around the United States.  Many of these methods have been used to analyze sentencing procedures (e.g., \cite{christin,kehl}) and parole decisions and recidivism (e.g., \cite{dressel, johndrow}).  Less attention has been paid to algorithms that are used to securitize prison populations; algorithms we call \textit{carceral} algorithms.  A potential reason for this is that gaining access to the necessary data is more challenging as the data are not available in public records.    One such algorithm is the Pennsylvania Additive Classification Tool (PACT).  

The PACT is an algorithm implemented by the Pennsylvania Department of Corrections (PADOC) in 1991.  It generates a classification number to determine the custody level of an incarcerated person.  The PACT is meant to standardize the custody assignment process and to systematically sort prisoners based on a prediction of their institutional behavior. The PACT number equates to a custody level: CL–1 (community corrections), CL–2 (minimum), CL–3 (medium), CL–4 (close), and CL–5 (maximum). These custody levels have dramatic consequences for incarcerated people as they will determine much of their experience in prison.  A claim that has accompanied the PACT since its creation is that its use will reduce bias because it was meant to be objective and determined by an incarcerated person's behavior \cite[pp. 37--38]{hardyman}.  

\begin{figure}
    \centering
    \includegraphics[width=\textwidth]{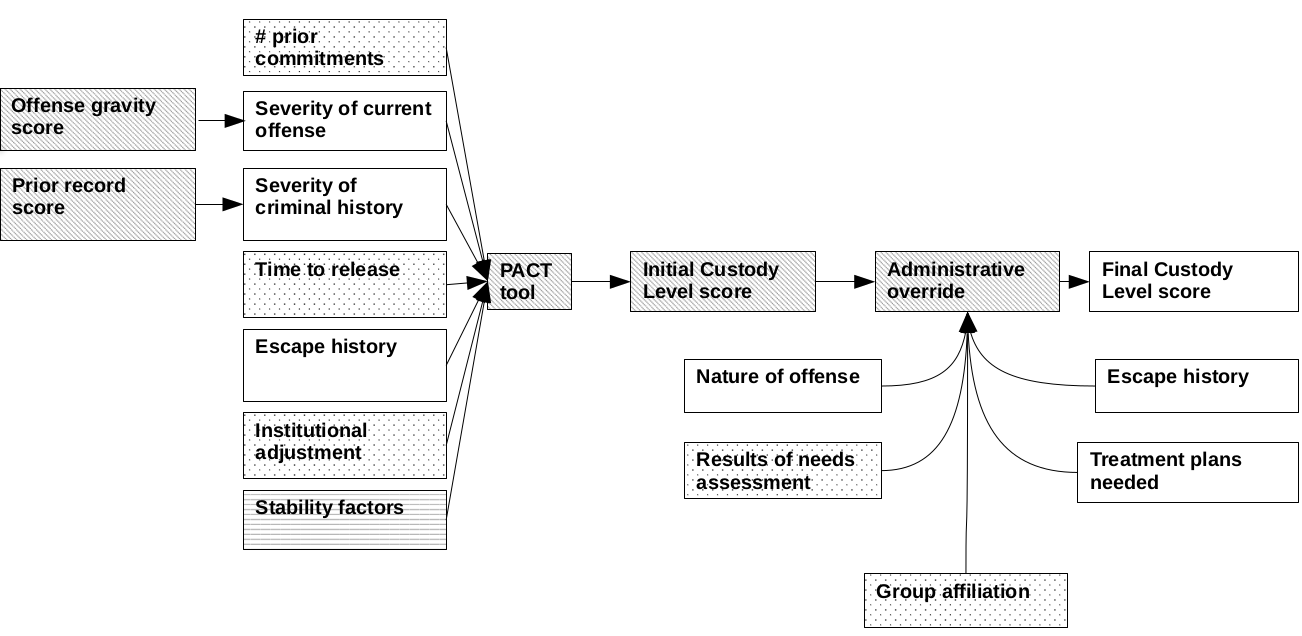}
    \caption{A schematic description of how an incarcerated person's custody level score is determined.  The process has two main parts:  a score determined by the PACT and then an override.  Variables in boxes with a background of diagonal lines indicate a lack of transparency about how that variable is calculated; variables in boxes with a background made of little dots are ones where the values are known to have a societal bias; the variable in a box whose background is horizontal lines represents several variables (e.g., marital status, employment status, etc.) that are somewhat arbitrary; and variables in white are ones that are either not known to be biased or whose method of calculation is known.}
    \label{fig:flow-chart}
\end{figure}

As can be seen in Figure~\ref{fig:flow-chart}, there are many variables in a two-stage process to determine an incarcerated person's initial custody level.  A score is determined by the PACT tool and then an override (discretionary or administrative) is applied to that score.  In this paper we recreate the algorithm and the override process using various logistic regression models.   We build the models from data we acquired from the PADOC and find that logistic regression replicates the PACT fairly well.  We interpret the model parameters to describe biases that are reinforced by the use of the PACT and discuss the role that the override process plays in the decision.

The paper is organized as follows.  In the next section we provide background related to biases in criminal justice algorithms and the PACT.  In the subsequent section we describe our methods and then we state our main results.  We end with some discussion and future directions of study.

\subsection*{Main contribution}  In this article we carry out several statistical analyses of the PACT, identify the most important features in the models and use our results to quantify biases in the procedures.  Because the algorithm is redacted and only applied to incarcerated people, few analyses of this kind have been conducted.  In addition to recreating the PACT in a statistically rigorous way, we also begin to study the override process and how it interacts with the initial score calculated by the PACT.  

\section{Background}
The Pennsylvania Additive Classification Tool (PACT) is a tool used by the Pennsylvania Department of Corrections (PADOC) to assign an incarcerated person a custody level.  The PACT calculates a score and then the prison in which they are being incarcerated has the option to override the score and determine a person's final custody level.  The PACT is confidential and redacted from Pennsylvania statute, but in \cite{dmmr1} we closely analyze the documentation provided by the PADOC \cite{PACT} to determine which variables are being used in the PACT.  The PACT process is summarized in Figure~\ref{fig:flow-chart}.  The PACT is used on incarcerated people and such algorithms have not been studied very extensively, unlike sentencing algorithms and parole/recidivism algorithms.  

A well-known algorithm that has been studied extensively (and not uncontroversially) is the Correctional Offender Management Profiling for Alternative Sanctions (COMPAS).  COMPAS is a collection of tools to manage data and to inform decisions related to defendants and parolees recidivating.  In 2016 ProPublica published an article \cite{propublica1} in which they argue that the application of COMPAS is biased against Black people; their methodology is described in \cite{propublica2}.  The COMPAS tool is complicated and hard to interpret. In fact, it has been shown that COMPAS has an accuracy of 0.65 while groups of individuals with limited or no expertise in criminal justice has an accuracy of 0.67, when their responses are pooled \cite{atlantic}; researchers have created optimal sparse rule lists that are approximately as accurate as the COMPAS \cite{angelino}; and others \cite{gsell} show that the number of prior convictions predicts recidivism about as accurately as COMPAS.  

The PACT, by virtue of being confidential and redacted, is also hard to interpret.  Also, by virtue of being a carceral algorithm and therefore less visible in its impact, there is an even greater need for understanding the procedure by which people are assigned custody levels.  The custody level a person is assigned has been shown to have an impact on the person's well-being and likelihood of being paroled.  For example, in \cite{visitation}, it is shown that a person having visitations while incarcerated is correlated with the person being released and the granting of visitations in Pennsylvania is related to the person's custody level (e.g., according to \cite{visitorguide}, ``An inmate in the Restricted Housing Unit under Disciplinary Custody may receive 1 video visit per month (Monday through Friday).'')  As another example, custody level has been studied as it relates to conduct violations (e.g., \cite{violations} shows a strong positive correlation between custody level and the number of conduct violations a person is given) and how it relates to an incarcerated person's well-being (e.g., \cite{violations} shows a correlation between program participation, visitation, etc. and a person's adjustment to being incarcerated and those factors are dependent on a person's custody level).

In \cite{dmmr1}, we have studied the PACT from a historical perspective and argue that it has competing sets of goals.  One set of goals includes a need to better securitize the incarcerated population and to deal with that population in a more efficient way.  Another set of goals includes improved monitoring of what an incarcerated person is doing in terms of programs, physical and mental health, etc.  Two main algorithmic findings of \cite{dmmr1} were (1) identifying the variables used in the PACT and matching them with the data dictionary provided to us by the PADOC and (2) a quantitative description of how much data is missing about incarcerated people and the low quality of the data that there is.  In \cite{dmmr2}, we developed a random forest model to understand the PACT and, using this model, we determined the long term impact of reapplying the PACT to incarcerated people; this was an attempt at modeling the reclassification process described above and its long term behavior.

\section{Methods}  The goal of our models is to predict the custody level of an incarcerated person based on the variables identified in Figure~\ref{fig:flow-chart}.  We do this in two stages (PACT classification and overrides) and model the process using logistic regression.  Recall that a PACT score and a custody level score are both numbers from 1 to 5.  We build three different models:  a multinomial model that predicts scores from 2 to 5 (we omit scores of 1 because those incarcerated people are on parole and not in a prison); a binary model that predicts maximum security (level 5) versus custody levels 2, 3 or 4; and a binary model that predicts high custody levels (levels 4 or 5) versus low custody levels (levels 2 or 3).  For convenience we will refer to these classification tasks respectively as multinomial, max vs regular and high vs low.  The code that we used to generate the tables, figures and statistics in this paper (but not the data as it is sensitive and has not been de-identified) is available at \cite{repo}.

\subsection{Preparing the data}

The data we received from the PADOC covered the years 1997, 2002, 2007, 2012, and 2017 and include both people who were on parole and those who were not.  The variables we requested were those that were related to parole decisions. We received data on more than 280,000 distinct incarcerated people, but of those only 146,793 were incarcerated in the years we requested.

\begin{figure}
    \centering
    \includegraphics[width=\textwidth]{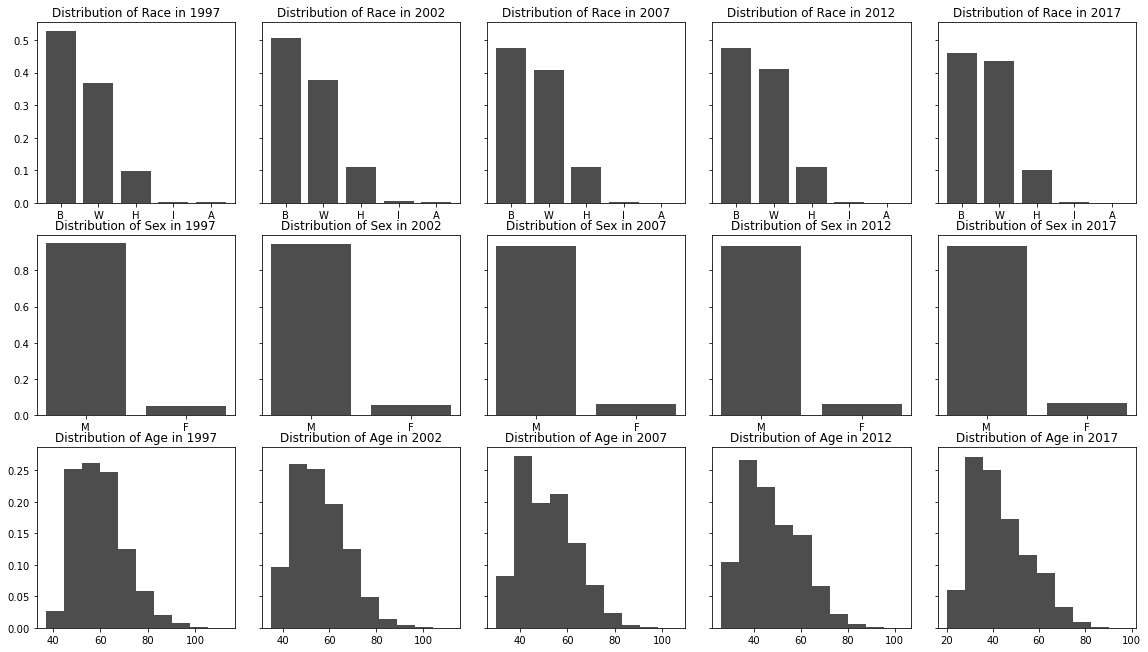}
    \caption{Basic demographic data for incarcerated people in the PADOC system in the years 1997, 2002, 2007, 2012, 2017}
    \label{fig:demographics}
\end{figure}

In Figure~\ref{fig:demographics}, we see a sequence of graphs that show the racial demographics, sex demographics and age distribution of the people in the original data pull.  Note that recently the population has trended younger, the ratio of the number of male incarcerated people to the number of female incarcerated people has remained stable and the ratio of the number of Black incarcerated people to the number of Whites has changed from roughly 3:2 in 1997 to approximately 1:1 in 2017. This disparity remains in spite of Black people making up only about 11\% of Pennsylvania’s general population. In this study we only analyze the data for those people incarcerated in 2017.

Two of the inputs for the PACT are the severity of an incarcerated person's current offense and their criminal history.  In the data we were given the abbreviations for the violations the incarcerated people were convicted of, but an actual gravity record score or prior offense score were missing.  So, in order to get meaningful data, we merged the data in our data set with the information in the Pennsylvania Criminal Code \cite{criminalcode} to assign a prior record score and an offense gravity score to each person for whom we had an abbreviation of their violation(s).  When there was not an exact match between the abbreviation and the Criminal Code, we assigned a range to the person.  For example, criminal code 3926(b) corresponds to ``Theft of services-divert service" and, depending on the amount stolen, this carries an offense gravity score from 1 to 3 points.  In our data, in the absence of more specificity than the criminal code number, we use the maximum of each person's offense gravity scores and prior record scores. 

Among the variables for which we have data, we find 5 variables related to whether or not an incarcerated person has ``escaped''.  An escape is defined in Title 18 Section 5121 \cite{escape} to occur when ``A person\dots unlawfully removes himself from official detention or fails to return to official detention following temporary leave granted for a specific purpose or limited period'' and where ``official detention'' means ``arrest, detention in any facility for custody of persons under charge or conviction of crime or alleged or found to be delinquent, detention for extradition or deportation, or any other detention for law enforcement purposes; but the phrase does not include supervision of probation or parole, or constraint incidental to release on bail.''  Since we are interested in whether a person has ever escaped or is an habitual escaper, we only keep the variable that shows if they have ever escaped and the variable that shows if they have escaped at least 5 times.

See Table~\ref{tbl:vars} for descriptions of the variables we used in our models.

\begin{table}\small
\begin{tabular}{|p{1.4in}|p{3.1in}|}\hline
Variable & Description \\\hline
gender\_female & A binary variable indicating whether the person is identified as female (1) or male (0)	
\\\hline
age\_gt\_45& A binary variable indicating whether the person is older than 45 years old (1) or not  (0)	
\\\hline
age\_lt\_25& A binary variable indicating whether the person is younger than 25 years old (1) or not  (0)
\\\hline
race\_B	&  A binary variable indicating whether the person is identified as Black (1) or not (0)
\\\hline
race\_A	&  A binary variable indicating whether the person is identified as Asian of Pacific Islander (1) or not (0)
\\\hline
race\_H	&  A binary variable indicating whether the person is identified as Hispanic (1) or not (0)
\\\hline
race\_I	&  A binary variable indicating whether the person is identified as American Indian (1) or not (0)
\\\hline
race\_O	&  A binary variable indicating whether the person is identified as belonging to a Nonwhite race or ethnicity other than Black, Asian, Hispanic or American Indian (1) or not (0)
\\\hline
off\_1\_prs\_max& A quantitative variable representing a person's maximum prior record score 
\\\hline
off\_1\_gs\_max& A quantitative variable representing the maximum gravity score a person could have received 
\\\hline
prior\_commits & The number of times the person has been previously committed to the PADOC 
\\\hline
ic\_institut\_adj & A person's institutional adjustment score used during initial classification 
\\\hline
re\_discip\_reports& The number of disciplinary reports a person was given since the previous reclassification 
\\\hline
escape\_hist\_1&	 A binary variable representing whether there was an escape as described in the text 
\\\hline
escape\_hist\_5&	A binary variable representing whether there was a fifth escape as described in the text
\\\hline
mrt\_stat\_DIV & A binary variable representing whether the person was divorced at the time of commitment (1) or not (0)	
\\\hline
mrt\_stat\_SEP& A binary variable representing whether the person was separated at the time of commitment (1) or not (0)	
\\\hline
mrt\_stat\_MAR& A binary variable representing whether the person was married at the time of commitment (1) or not (0)	
\\\hline
mrt\_stat\_WID& A binary variable representing whether the person was widowed at the time of commitment (1) or not (0)	
\\\hline
employed & A binary variable representing whether the person was employed at the time of commitment (1) or not (0)	
\\\hline
problematic\_offenses & A binary variable representing whether the person's offense was ``problematic'' as described in the text
\\\hline
problematic\_conditions &  A quantitative variable that counts how many of the following apply to an incarcerated person: tendency to escape, a problem with alcohol, a problem with drugs, suicidal thoughts, in psychological treatment, committed a crime that was sexual in nature, and is aggressive with other incarcerated people
\\\hline
\end{tabular}
\caption{A summary of the variables used in the construction of our models.}\label{tbl:vars}
\end{table}

\subsection{Logistic regression}

As mentioned above, we model four different classification procedures using logistic regression:
\begin{enumerate}
\item we use multinomial logistic regression to classify an incarcerated person into custody levels 2 through 5 (recall that we dropped custody level 1 because those individuals, while incarcerated due to being on parole, are not housed in a prison);
\item we use a binary logistic regression to classify a person into maximum security (custody level 5) or not;
\item we use a binary logistic regression to classify a person into high security (custody levels 4 or 5) or not;
\item we use a binary logistic regression to classify whether a person is receiving an override to a higher custody level or not.
\end{enumerate}
We model these four processes both for the data related to a person's initial classification and for the data related to a person's reclassification. 

\begin{figure}
\includegraphics[width=.3\textwidth]{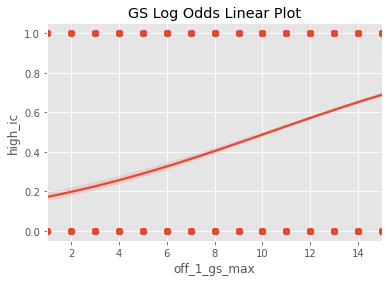} 
\hfill
\includegraphics[width=.3\textwidth]{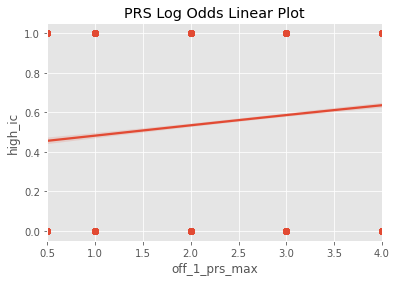}
\hfill
\includegraphics[width=.3\textwidth]{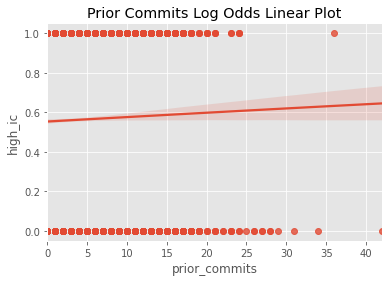}
\caption{Plots of the log odds of being assigned a high custody level versus, respectively, from left to right, a person's gravity score, a person's prior record score and a person's number of prior commitments.  }\label{fig:lin-log-odds}
\end{figure}

\begin{figure}
\includegraphics[width=.45\textwidth]{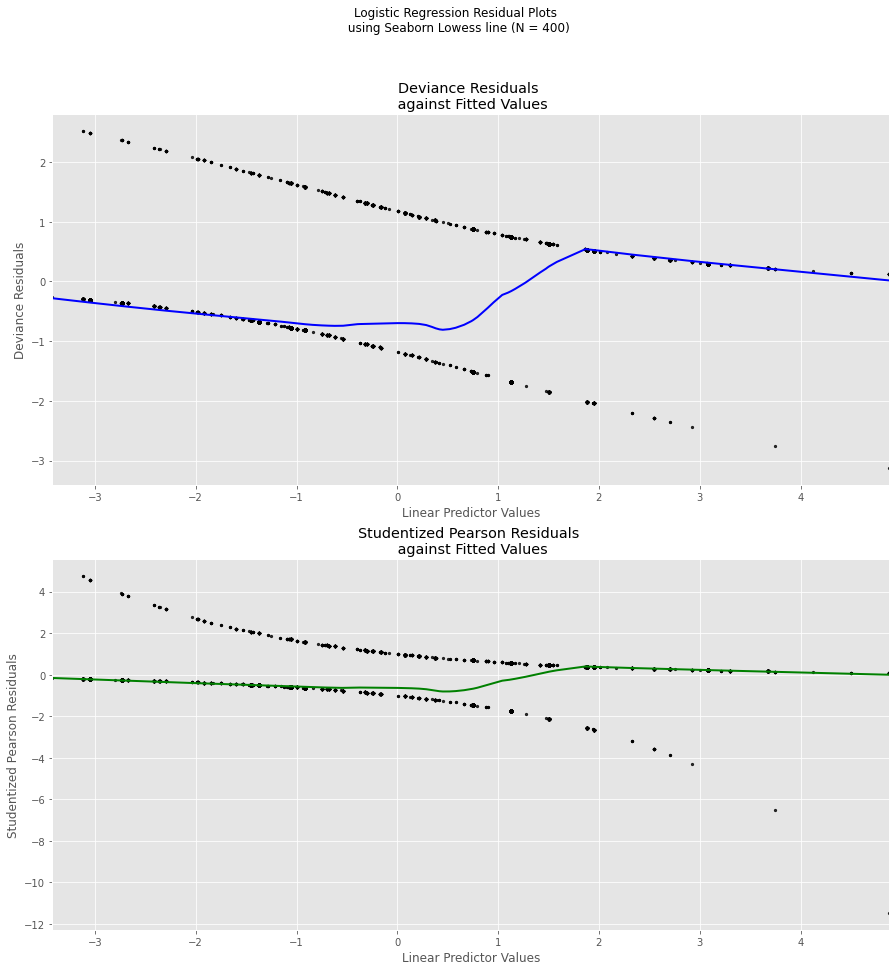}
\hfill
\includegraphics[width=.45\textwidth]{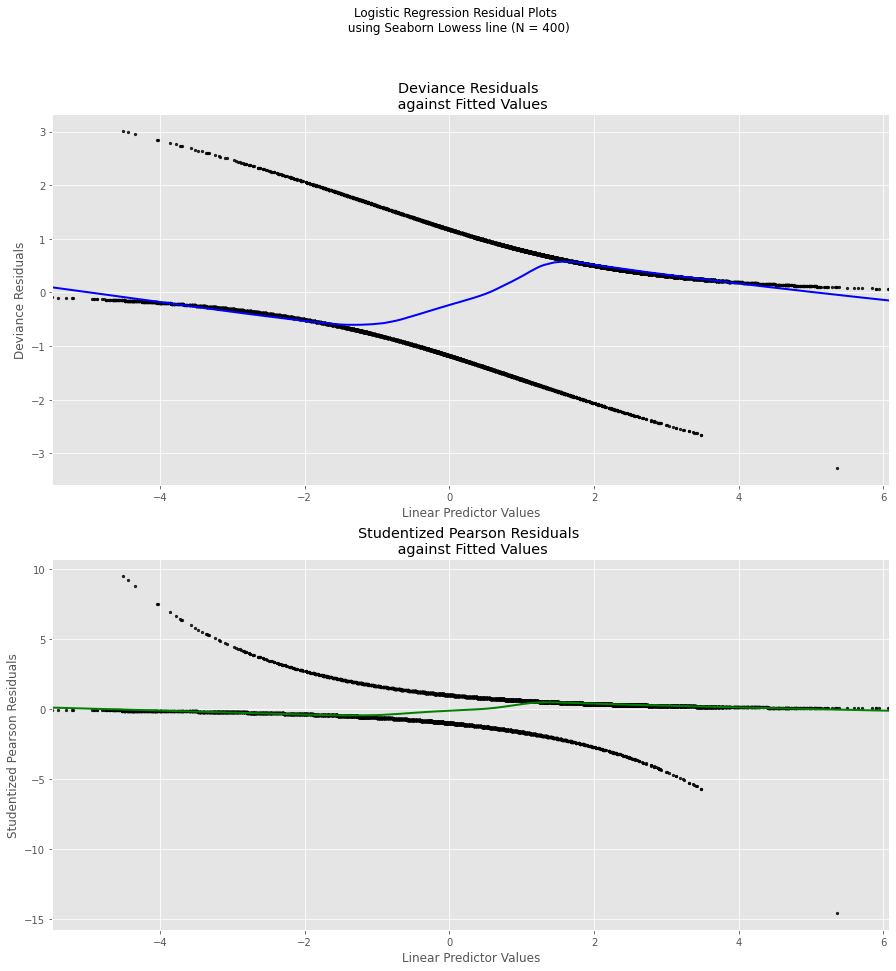}
\caption{Plots of the deviance and studentized Pearson residuals against fitted values for the initial classification after feature selection (left) and after the subsequent addition of demographics variables (right).}\label{fig:resids}
\end{figure}

\begin{figure}
    \centering
    \includegraphics[width=0.8\textwidth]{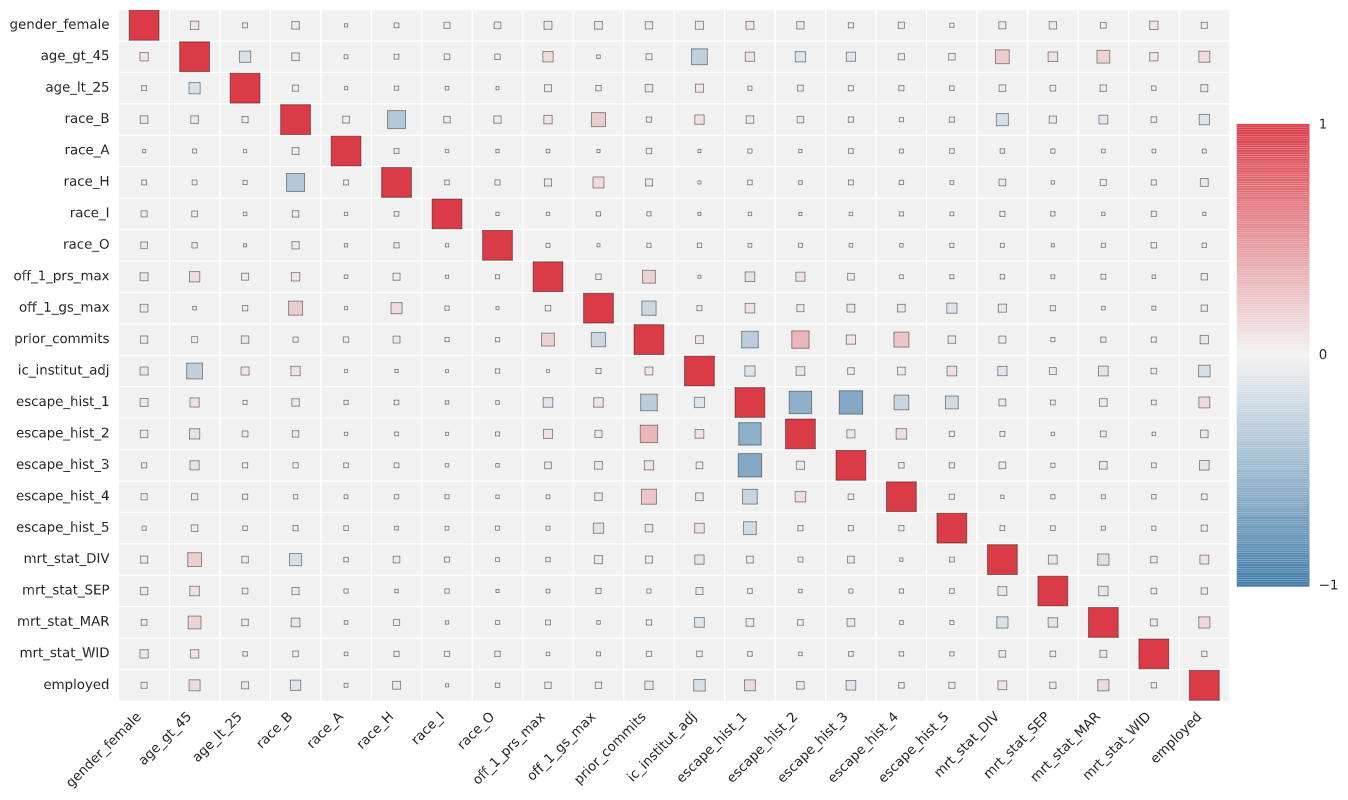}
    \caption{Correlation matrix of variables used in the PACT tool.  The area of the squares are proportional to the correlation coefficient. }
    \label{fig:logit-ic-corr}
\end{figure}

For every model that we develop, we check the following assumptions.  First, we see if the classes are balanced.  For example, in the case of classifying a person into maximum security, the two classes are very imbalanced.  In cases like this we used the SMOTE algorithm as described in \cite{smote} and as implemented in the Imbalanced-learn Python package \cite{imblearn}.  We had to balance the classes in most cases.  Second, we checked for a lack of collinearity among our independent variables by considering  the pairwise correlation coefficients.  In all cases the correlation coefficients were between $-0.3$ and $0.3$.  Third, we checked whether the continuous independent variables (the number of prior commits, the number of disciplinary reports, the gravity score, the prior record score) in each model were linearly related to the log odds.  In all cases, the regression plot had the desired shape.  Fourth, following \cite{kutner} we plot the deviance and studentized Pearson residuals versus the fitted values and look for a flat line with a $y$-intercept of zero.  In all cases, the studentized Pearson residuals had the expected shape and in the majority of cases so did the plot of deviance residuals.  See our code at \cite{repo} to support these claims and see Figures~\ref{fig:lin-log-odds}, \ref{fig:resids}, and \ref{fig:logit-ic-corr} for examples.

After confirming that the model satisfied the expected conditions, we conducted recursive feature selection to determined which features had the most influence in the classification.

\section{Results}









Now, we describe the various logistic models we fit to the data.  As mentioned above, in all cases we checked that the data satisfied the necessary conditions, oversampling using SMOTE if the classes were imbalanced.  More details and more statistics can be found in the notebooks at \cite{repo}.

\subsection{Classification into high versus low security}

We describe our results for our modeling of the initial classification of incarcerated people into high security (custody levels of 4 or 5) or into low security (custody levels 2 or 3).  After preparing the data set there was complete data for $N=13185$ incarcerated people, of whom $7750$ were in the higher custody level.  We developed a logistic regression model using the relevant variables in Table~\ref{tbl:vars}, and highlight the following results.  First, a person's institutional adjustment (2.09 times more likely to be classified to the higher custody level for every point higher a person scores) and whether they are habitual escapees (a person with at least 5 attempted escapes is 4.01 times more likely to be classified to the higher custody level than a person who does not have at least 5 attempted escapes) affect their likelihood of being assigned a higher custody level.  These are both significant at the significance level $\alpha = 0.05$ and indicate how the PACT is used to control the incarcerated population.  Second, in terms of biases, we find that a Black person is $1.27$ times more likely to be classified to high custody level than a person who is not Black, a person younger than 25 is $1.47$ times more likely than a person who is not, a person older than 45 is $0.28$ times more likely to be classified to a higher custody level, and a person identified as female is $0.34$ times more likely than a person not identified as female.  We report that this model has a precision of 0.83, a recall of 0.83 and an $F_1$-score of 0.83 (all of these are calculated as weighted averages).  Finally, we observe that the most important features according to this logistic regression model were whether or not the person is female or not, whether they were Black or not, whether they were Asian or Pacific Islander or not and whether they were Hispanic or not.

A similar analysis for our model for reclassification was done, with the following results. After preparing the dataset there was complete data for $N=14572$ incarcerated people, of whom $2547$ were assigned to the higher custody level.  Since the classes were so imbalanced, we used SMOTE to oversample and developed a logistic regression model for which we highlight the following results.  First, the number of disciplinary reports a person receives affects the probability of them being assigned to a higher custody level (for every additional disciplinary report a person receives, they are 3.49 times more likely to be assigned to a higher custody level).  Second, in terms of biases, a person identified as female is $0.09$ times as likely to be assigned to a high custody level than someone who is not and someone who is Asian or Pacific Islander is $0.09$ as likely.  All of these were significant at the $\alpha=0.05$ level.  The recall, precision and $F_1$-score of this model were all $0.87$, again reported as a weighted average.  Finally, the most important features in this model are whether the person is identified as female or not; whether or not the person was older than 45, whether or not the person was widowed or whether or not the person was employed.

\subsection{Classification into maximum vs regular security} 

Our results related to the modeling of when a person gets placed into maximum security upon initial classification are as follows.  There were $N=13815$ incarcerated people for which we had complete data and of those $51$ were in maximum security custody level 5.  These are extremely imbalanced classes and so we used SMOTE to balance the classes by oversampling.  Also, in order to avoid complete separation, we had to remove several indicator variables (namely, the ones about whether or not the incarcerated person is Asian or Pacific Islander, American Indian, or a member of the race other than those listed; as well as the ones related to whether the person was divorced, separated or widowed). After making these adjustments to the model, we first found that for every point higher a person's institutional adjustment score is, they are $1.40$ times more likely to be in maximum security and we found that for every point higher their offense's gravity score is they are $1.09$ times more likely to be in maximum security.  Second, in terms of demographic features, we find that a person who is younger than 25 is $1.22$ more times more likely to be placed into maximum security and a person who is identified as female is $0.004$ times as likely to be placed in maximum security.  All of these were significant at the $\alpha=0.05$ level.  The recall, precision, and $F_1$-score each had a reported weighted average of $0.83$.  Finally, the most important features according to the model were if the person ever escaped, if they escaped 5 or more times, whether they were employed when they entered the system and whether they were younger than 25.

The same analyses were carried out for whether a person is placed in maximum security upon reclassification.  There were $N=14572$ incarcerated for which we had complete data related to classification into maximum security and of those $763$ were in maximum security.  The classes were imbalanced and so we oversampled using SMOTE and we did not have separation and so we did not need to remove any variables.  First, we note that for every extra disciplinary report a person had, they were $2.09$ times more likely to be in maximum security.  Second, we note that if the person is older than 45 they are $0.16$ times more likely to be placed into maximum security and if the person is identified as female, they are $0.03$ times more likely to be placed into maximum security.  These are significant at the $\alpha=0.05$ level.  We report that the recall, precision and $F_1$-score were each equal to 0.81.  Finally, we observe that the variables that were most important were whether the person was older than 45, the number of disciplinary reports, the person's identified gender and whether or not they were identified as Asian.

\subsection{Multinomial regression}

In this section we summarize the results from the multinomial regression into custody levels 2 through 5.  We start with our results from our model of the initial classification.  There were $N=13815$ people for whom he had complete data, 1324 people in custody level 2, 4741 in custody level 3, 7699 in custody level 4 and 51 in custody level 5.  Because of the class imbalance, we used SMOTE to oversample.  As we are focused on biases relative to demographic features, we only report those coefficients related to a person's identified race, gender and age.  For custody level 2, we find that being identified as a women makes one $2.13$ times more likely to be in custody level 2, being older than 45 make one $4.40$ times more likely to be in custody level 2, being younger than 25 or being nonwhite makes you less likely to be in custody level 2.  In custody level 3, we find that being identified as a woman makes it $1.56$ times more likely that person will be in custody level 3, a person older than 45 is $1.60$ times more likely to be in custody level 3, a person younger than 25 is $1.05$ times more likely to be in custody level 3, and people who are, respectively, identified as Black, Asian or Pacific Island, Hispanic, American Indian or Other are $1.22$, $0.89$, $1.46$, $1.05$, and $1.08$ times more likely to have been placed in custody level 3.  In custody level 4, a person identified as a woman is $1.88$ times more likely to be in custody level 4,  a person older than 45 is $0.59$ times more likely to be in custody level 3, a person younger than 25 is $3.08$ times more likely to be in custody level 4, and people who are, respectively, identified as Black, Asian or Pacific Island, Hispanic, American Indian or Other are $2.79$, $3.93$, $2.39$, $2.03$, and $2.44$ times more likely to have been placed in custody level 4.  In custody level 5, the relative odds are increased by a factor of $3.08$ if the person is under 25, but the person's institutional adjustment score is really driving the classification into custody level 5.  The most important features in the model are the person's marital history and whether they have escaped at least 5 times.  The accuracy of this model was $0.722$ (an average of a 10-fold repeated stratification) before balancing the classes and $0.678$ after balancing the classes.

Now we turn our focus to reclassification.  In  custody level 2, we find that being identified as a women makes one $5.78$ times more likely to be in custody level 2, being older than 45 make one $4.37$ times more likely to be in custody level 2, being younger than 25 makes it $1.11$ times more likely to be in custody level 2, and people who are, respectively, identified as Black, Asian or Pacific Island, Hispanic, American Indian or Other are $1.39$, $1.27$, $2.31$, $0.69$, and $1.90$ times more likely to have been placed in custody level 2.  In custody level 3, we find that being identified as a woman makes it $1.81$ times more likely that person will be in custody level 3, a person older than 45 is $1.50$ times more likely to be in custody level 3, a person younger than 25 is $0.67$ times more likely to be in custody level 3, and people who are, respectively, identified as Black, Asian or Pacific Island, Hispanic, American Indian or Other are $0.95$, $0.81$, $0.93$, $1.97$, and $1.04$ times more likely to have been placed in custody level 3.  In custody level 4, a person identified as a woman is $0.35$ times more likely to be in custody level 4,  a person older than 45 is $0.46$ times more likely to be in custody level 3, a person younger than 25 is $1.56$ times more likely to be in custody level 4, and people who are, respectively, identified as Black, Asian or Pacific Island, Hispanic, American Indian or Other are $1.16$, $1.00$, $0.91$, $0.90$, and $0.98$ times more likely to have been placed in custody level 4.  In custody level 5, the relative odds are increased by a factor of $2.35$ if the person is widowed (an artifact of the small sample size) or by a factor of $2.06$ for every additional disciplinary report they receive.  The accuracy of this model was $0.776$ (an average of a 10-fold repeated stratification) before balancing the classes and $0.573$ after balancing the classes.

\subsection{Overrides}  

The override procedure is even less legible than the PACT procedure since it is, by its very nature, less explicit, as overrides are given for both discretionary and administrative reasons.  In fact, an override is defined in \cite{PACT} to be ``a change in the PACT generated custody level based upon aggravating or mitigating factors that are not considered on the PACT instrument'' and an administrative override is defined to be ``a factor applied against an inmate’s scored custody level that is mandatory in nature emanating from law or agency policy and must be applied where appropriate.''

We predict whether someone receives an override to a higher custody levels by incorporating some of the variables identified in Figure~\ref{fig:flow-chart}:  for example any data about group affiliations (e.g., membership in gangs), nature of the offense (e.g., was it sexual in nature), and about treatment plans needed by the incarcerated person.  

Data about group affiliation is no longer collected by the PADOC and so we focus on data related to the nature of the offense and the treatment programs needed by the incarcerated person.  According to \cite{problematic}, an offense is determined to be ``problematic'' as follows:
\begin{quote}
The following offenses shall be considered problematic:
\begin{enumerate}
\item[a.] criminal homicide;
\item[b.] murder 1, 2, or 3;
\item[c.] voluntary manslaughter;
\item[d.] manslaughter of law enforcement officer;
\item[e.] 3rd degree murder of unborn child;
\item[f.] kidnapping;
\item[g.] any sex offense (with the exception of prostitution); and/or
\item[h.] attempt/solicitation/conspiracy to commit any of the above offenses. 
\end{enumerate}
\end{quote}
A variable in the data set we received from the PADOC tells us if the nature of the offense is ``problematic'' or not and we use that variable in the model.  

In order to determine the treatment needs of an incarcerated person, we consider variables related to whether or not the person has a verified history of drug use, alcohol use, assaultive behavior, sexual offenses, mental illness, escapes and suicidal ideation.  We count the number of such histories a person has and make a new variable representing that count.

For initial classification we have $N=13816$ and with 253 receiving an override to higher custody level; like usual, we use SMOTE to balance the classes.  We find that the most important features in the model are whether or not the person is identified as female, whether they are American Indian, whether their race or ethnicity would be classified as Other and whether they are widowed.  We find that for a person whose offense was labeled as problematic, they are $3.54$ times more likely to receive and override to a higher custody level and we find that someone who has escaped at least 5 times is $3.86$ times more likely to receive an override to a higher custody level.  A person identified as female is $0.62$ times as likely to receive an override at a custody level and this is significant at the $\alpha=0.05$ level.  The other odds ratios for demographic features were not significant and so we do not report them here.  This model performs less well overall and struggles to predict those who have receive an override to a higher custody level.  The weighted average of the precision is $0.97$, of the recall is $0.76$ and of the $F_1$-score is $0.85$ but if we restrict to predicting those who receive an override to a higher custody level the numbers are, respectively, $0.03$, $0.44$ and $0.06$.

For reclassification we have $N=16543$ and with 2797 receiving an override to higher custody level; yet again, we use SMOTE to balance the classes.  We find that the most important features in the model are whether or not the person is identified as female, is younger than 25, has problematic offenses and whose marital status is that they are separated.  We find that for a person whose offense was labeled as problematic, they are $3.43$ times more likely to receive and override to a higher custody level and we find that someone who has escaped at least 5 times is $2.71$ times more likely to receive an override to a higher custody level.  A person identified as female is $0.45$ times as likely to receive an override at a custody level, a person who is Black $1.23$ times more likely, a person who is Hispanic is $0.80$ times more likely, a person older than 45 is $1.59$ times more likely, a person younger than 25 is $0.27$ times more likely; these are significant at the $\alpha=0.05$ level.  This reclassification model performs better than the initial classification model:  its weighted average precision, recall and $F_1$-scores are, respectively, $0.87$, $0.81$ and $0.83$ but those same statistics for predicting whether someone receives a higher custody level are, again respectively, $0.48$, $0.80$ and $0.60$.

\section{Conclusion and discussion}

We conducted an analysis of the processes by which people incarcerated in Pennsylvania are assigned custody levels.  We built logistic regression models for each of the following processes:  (1) the application of the PACT to determine a person's initial custody level, (2) the application of the PACT to determine if someone should be placed in maximum security , (3) the application of the PACT to determine whether someone should be placed in high security level, (4) the process by which overrides to initial classification are made, and (5-8) the same processes but for reclassification.

Of the various models for initial classification, we focus on the one that classified people into custody level 4 or 5 as opposed to custody level 2 or 3.  There are various reasons we feel justified in this decision.  First, there are very few individuals in custody level 5 (51 out of more than 10000 in total), and so models that attempt to predict custody level 5 are going to have a hard time doing so.  Second, thinking of custody levels 4 and 5 as ``high'' security is sensible because the definition of custody level 4 in \cite{PACT} reads that incarcerated people at the custody level require  ``continuous direct and indirect supervision.''  For custody levels 2 and 3, on the other hand, the definitions in \cite{PACT} read that, in custody level 2, incarcerated people require ``only intermittent, direct observation by staff'' and, in custody level 3, incarcerated people ``require frequent, direct supervision.''  The level of securitization is qualitatively different between custody levels 3 and 4 but similar at custody levels 2 and 3 and so we group them accordingly.  Third, the models performed the best.

The initial classification process of incarcerated people into low or high custody is, according to our model, driven the most by protected traits of the individual, namely, whether or not the person was identified as female, or as Black or as Hispanic.  Moreover, people identified as Black are much more likely to be placed in a high custody level than those who are not identified as Black.  This ratio would likely be even higher if we compared Black and White inmates as opposed to Black vs Nonblack inmates.  An incarcerated person identified as female is much less likely to be placed in a high custody level than those who are not identified as female.  A young person is much more likely to be placed into a high custody level and an older person much less.  There is an institutional adjustment score with a high odds ratio:  for every one point increase in this score, a person is more than twice as likely to be placed into a higher custody level.  However, the calculation of this score is mysterious in the sense that there is no way for one to know what goes into that score.  A person's escape history also has an impact, especially whether they have at least 5 times:  such a person, if 4 times more likely to be placed into a high custody level.  But very few incarcerated people have escaped 5 or more times, and so this is not a driving force behind the predictions.  So, among the features that are interpretable, the features that carry the most weight are features that should be protected and the features that are protected exhibit a great deal of bias.

Our model for the reclassification process of people into a low or high security level is qualitatively different.  This is due to the fact that a person's custody tends to lower over time and, additionally, as detailed in \cite[pg. 49]{PACT}, a person about whom not much is known is placed into custody level 4 (``Newly received inmates who are unclassified are assigned to this level'').  Whether or not a person has been identified as female and whether the person is older than 45 are two of the features that carry the most weight and they are quite biased:  a person who is identified as female is ten times less likely to be reclassified into a high custody level and people who are older than 45 are almost 4 times less likely to be reclassified into a higher custody level.  For every disciplinary report a person receives, they are almost 3.5 times more likely to placed into a high custody level.  While this might appear to be evidence that the system is working, it is important to remember that there is reason to believe that the process of being given disciplinary reports is biased against Black incarcerated people.

Our conclusions related to our model of the override process are less conclusive, in part because those models are less accurate.

One thing that caught our attention was how in none of the models neither the gravity score of a person's offense or their prior record nor the number of prior commitments the person had appeared to have much impact on a person's custody level.  Not only is this borne about by how close their odds ratios are to one, but also by how close the means were to each other when calculated for each group (the mean number of prior commits was $3.22$ for those in the low custody level and $3.33$ for those in the high custody level; the means for prior record score were, respectively, $2.36$ versus $2.65$; the means for gravity score were, respectively, $10.9$ versus $12.5$).

There are several next steps we will pursue in future work, in addition to modeling the institutional adjustment score and the number of disciplinary reports used in reclassification.  The initial goal of the project was to determine what things influence parole decisions in Pennsylvania.  The PACT is one such thing and as a result of this paper, we now have a better understanding of it.  Also, we want to continue exploring the override process in order to better understand how it works.  For example, is it possible to determine whether an override was administrative or discretionary?  

A limitation of our findings is the nature of the sample that we were provided by the PADOC.  On the one hand, we were given data for almost 300,000 people but because of the missingness of the data and the variables we used in our models, we could only build models based on smaller collections of data.  If there was any bias in the determination of people who have complete data, that bias will be transferred to the models we developed above.

\bibliographystyle{plain}
\bibliography{cj}
\end{document}